\pdfoutput=1 
\documentclass{article}
\usepackage{amsmath,epsfig}
\usepackage[preprint]{spconfa4}
\usepackage{amsfonts}
\usepackage{mathrsfs}
\usepackage{multirow}
\usepackage{amssymb}
\usepackage{color}
\usepackage{diagbox}
\renewcommand{\paragraph}[1]{\noindent\textbf{#1}}
\usepackage{hyperref}
\hypersetup{colorlinks=true,
            linkcolor=red,
            anchorcolor=red,
            citecolor=green}

\let\OLDthebibliography\thebibliography
\renewcommand\thebibliography[1]{
  \OLDthebibliography{#1}
  \setlength{\parskip}{0pt}
  \setlength{\itemsep}{0pt plus 0.3ex}
}

\begin{document}\sloppy

\def\x{{\mathbf x}}
\def\L{{\cal L}}

\title{A Multi-user Oriented Live Free-viewpoint Video Streaming System Based On View Interpolation}
%
\name{Jingchuan Hu\textsuperscript{1,2}, Shuai Guo\textsuperscript{1}, Kai Zhou\textsuperscript{1,2}, Yu Dong\textsuperscript{1,2}, Jun Xu\textsuperscript{1,2} and Li Song\textsuperscript{1,2,3}
}
\address{
\textsuperscript{1}Institute of Image Communication and Network Engineering, Shanghai Jiao Tong University\\
\textsuperscript{2}Cooperative Medianet Innovation Center, Shanghai Jiao Tong University\\ 
\textsuperscript{3}MoE Key Lab of Artificial Intelligence, AI Institute, Shanghai Jiao Tong University\\
Shanghai 200240, China\\ 
\{hujingchuan, shuaiguo, xxzk30, thesmallfish, xujunzz, song\_li\}@sjtu.edu.cn}

\maketitle

\begin{abstract}
As an important application form of immersive multimedia services, free-viewpoint video (FVV) enables users with great immersive experience by strong interaction. However, the computational complexity of virtual view synthesis algorithms poses a significant challenge to the real-time performance of an FVV system. Furthermore, the individuality of user interaction makes it difficult to serve multiple users simultaneously for a system with conventional architecture. In this paper, we novelly introduce a CNN-based view interpolation algorithm to synthesis dense virtual views in real time. Based on this, we also build an end-to-end live free-viewpoint system with a multi-user oriented streaming strategy. Our system can utilize a single edge server to serve multiple users at the same time without having to bring a large view synthesis load on the client side. We analysis the whole system and show that our approaches give the user a pleasant immersive experience, in terms of both visual quality and latency.

\end{abstract}
\section{Introduction}
\label{sec:intro}
The popularity of the 5G network and the development of edge computing technologies have brought huge changes to the video industry, with immersive and interactive media contents gradually becoming the future trend of media services. Free-viewpoint video(FVV) \cite{1167876}, as a representative of interactive and immersive media content, has recently received extensive attention and research. FVV allows users to interactively choose any direction and viewpoint to watch the scenario according to their own needs, without being restricted by the position of the camera used for shooting. This kind of user-autonomous interaction gets rid of the dependence on the director, thus giving the audience more immersion.

Unlike traditional Virtual Reality(VR), the “outside-in" interaction method of free-viewpoint video brings users a better immersive experience compared to the “inside-out" method of the former. FVV systems can be widely used in sports events, variety shows and other scenarios.

Typically, the server side of a FVV system includes a multi-view video acquisition module, a 3D spatial dynamic scene representation server and  viewpoint rendering(virtual viewpoint synthesis) servers as described in \cite{8247262}. Among them, virtual view synthesis technology has attracted growing attention from researchers and engineers recently.

Depth-Image-Based-Rendering (DIBR) technology is the most commonly used view synthesis algorithm for free-viewpoint systems \cite{8247262,  9431676, dong_elastic_2021}. In the DIBR process, pixels from the real viewpoint texture image are warped into the 3D space of the scenario, and then reprojected to the target viewpoint under the guidance of the depth map and camera pose to synthesize the virtual view. 

However, due to the dis-occlusion and black holes introduced in 3D image warping, the synthesis results are often unsatisfactory. Moreover, 3D image warping requires depth maps and camera poses for guidance, and it is time consuming to obtain depth maps using depth estimation methods \cite{schoenberger2016mvs, yao2018mvsnet}. Even though the coarse depth map can be obtained in real time using RGB-D cameras in some systems \cite{9431676}, the complexity of the DIBR algorithm makes it difficult to be deployed on low-cost clients. As such it can only be deployed on edge servers, thus DIBR-based systems cannot provide diverse interactions for multiple users at the same time.
\begin{figure*}
    \centering  
    \includegraphics[width=0.88\textwidth]{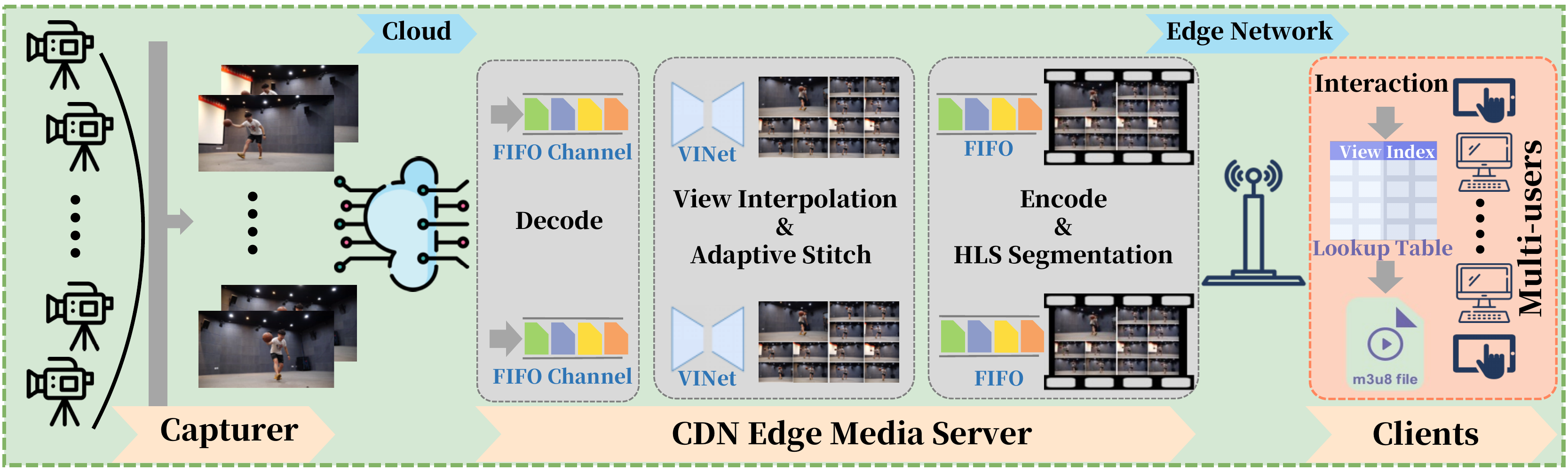}  
    \caption{Architecture of our multi-user oriented free viewpoint system.} 
    \label{System}
\end{figure*}

Some other view synthesis approaches, such as MPI \cite{zhou2018stereo, lin_deep_2021}, use multiple input images to construct a planar sweeping volume aligned to the novel viewpoint, and have that volume processed by a 3D CNN to generate an RGB$\alpha$ representation of the novel viewpoint. These neural scene representation methods have yielded some impressive results, but they are very GPU memory consuming and perform poorly in terms of runtime, making them challenging to be integrated into real-time FVV systems.

A variety of constraints have limited the widespread accessibility of FVV, thus there is an urgent need for a live free-viewpoint system that can serve multiple consumer-level terminals simultaneously. In this work, we propose a lightweight CNN-based view interpolation algorithm and a dense view streaming strategy for a real-time FVV system. By decoupling the server load from the amount of accessed clients, our live system can provide satisfactory free-viewpoint services to a theoretically arbitrary number of clients simultaneously without overburdening them.

Our key contributions can be summarized as follows:
\begin{itemize}
    \setlength{\itemsep}{0pt}
    \setlength{\parsep}{0pt}
    \setlength{\parskip}{0pt}
    \item We build an end-to-end multi-user oriented real-time FVV system that is capable of providing personalized interactions to multiple users at the same time. (Sec. \textcolor{red}{\ref{sec:system}})
    \item We propose an efficient lightweight view interpolation network called VINet, which can interpolate high quality intermediate virtual view in real time using two real captured views.  (Sec. \textcolor{red}{\ref{subsec:VINet}})
    \item For the generated dense view, we introduce a tile-based streaming strategy that can leverage a single edge media server to serve an arbitrary number of users simultaneously.  (Sec. \textcolor{red}{\ref{subsec:streaming}})
\end{itemize}

\section{System Overview}
\label{sec:system}
Generic FVV systems can be divided into centralized and distributed models. In the centralized model, the viewpoints required by different users are synthesised on the server side. Some existing real-time view synthesis algorithms require sufficient computing resources, thus only a limited number of user terminals can be served by one server. As the number of accessed users grows, the number of servers needs to be increased accordingly. This model has difficulty in managing high concurrency scenarios and incurs additional response latency during interactions.

The distributed model is capable of serving multiple users at the same time, since it performs the view synthesis process on the client side.  Nevertheless, the Multiview-Video-Plus-Depth (MVD) representation required for view synthesis needs to be transmitted to the users, which can incur a high transmission bandwidth. Additionally, the view synthesis algorithm requires significant processing power, which is unfriendly to some low-end user terminals.

The development of a real-time FVV system, capable of serving multiple users simultaneously and with low deployment cost, still poses a significant challenge that we have addressed in this work. We opt for a centralized model. However, our proposed VINet and dense view streaming strategy decouple the load of the edge media server from the number of clients, so that an arbitrary number of users can be served at the same time. Moreover, all modules on the server side execute in real time so that live media content can be produced.

As shown in Fig. \ref{System}, our complete end-to-end system consists of Capturers, Cloud, CDN Edge Media Sever, Edge Network and Clients. The Capturers acquire scene information from several different viewpoints. In our system, 12 cameras are arranged in a form of circular arc around the scene, with the angle (baseline) between each camera varying as the scenario changes. They are synchronised via hardware and are capable of capturing multi-view video at 4K/60FPS.

In some existing free-viewpoint video systems, depth sensors are often required to capture depth maps, which still face serious problems such as limited resolution and distance range for practical applications.  Furthermore,  these systems place high demands on the number of cameras and their arrangement, leading to high system costs. On the contrary, our system does not require depth maps and can achieve good results even with sparsely organized cameras, thereby satisfying the requirements of low cost and simplicity.

The CDN Edge Media Server utilizes its powerful computing resources to process the received video streams in real time and serves multiple clients. The kernel modules in the server include VINet and HTTP Live Streaming(HLS) based dense view streamer. VINet can generats dense viewpoints in real time. To satisfy the real time requirements, these adaptively stitched dense views will be encoded by NVENC\footnote{https://developer.nvidia.com/nvidia-video-codec-sdk} and then distributed by the streamer to the user terminals.

\begin{figure*}
    \centering  
    \includegraphics[width=0.85\textwidth]{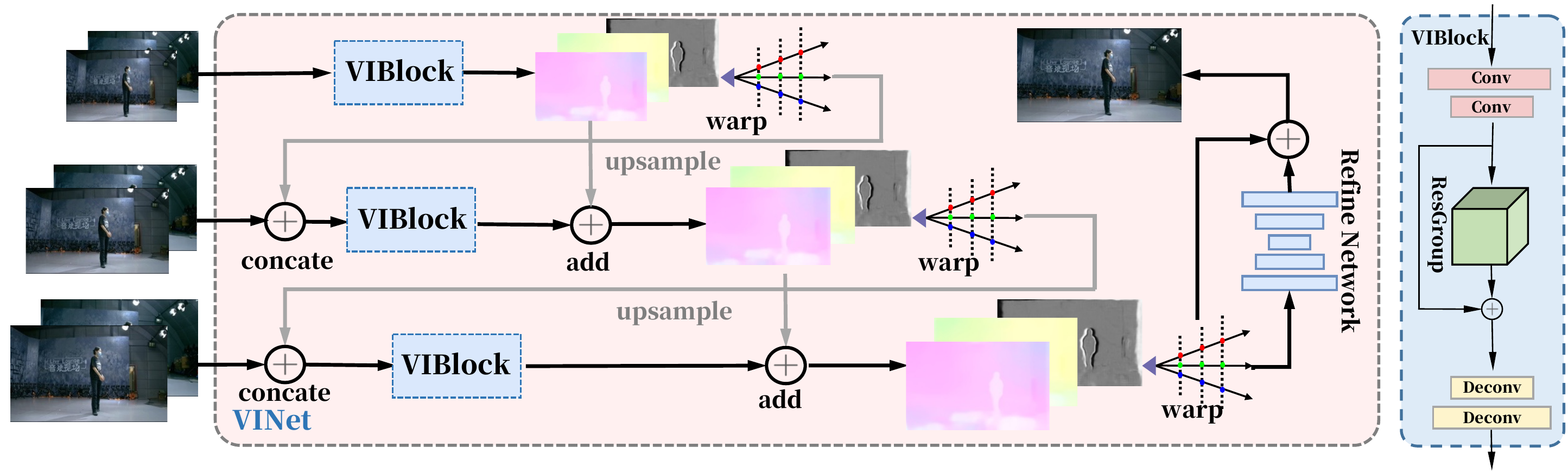}  
    \caption{Overview of our VINet. A total of three stages perform view interpolation in a coarse to fine manner, with each stage yielding a bidirectional flow and an occlusion-aware mask. An additional refine net is followed to enhance the visual quality.} 
    \label{VINet}
\end{figure*}

In our experimental configuration, multiple users are represented by a PC and a cell phone. It should be noted that although we use only two clients, an arbitrary number of clients can access our system and receive personalized interaction services. Since the client only needs to perform view extraction and decoding, which are tasks with minimal computational burden, most consumer-grade terminals can enjoy our free-viewpoint streaming service.





Additionally, to make our system more robust and smarter in task and resource management, we adopt the 4U resource model mentioned in \cite{dong_elastic_2021} to schedule our resources. 
Our VINet and multi-user oriented streaming strategy are discussed in Sec. \textcolor{red}{\ref{sec:method}}. As for some configuration details about our end-to-end system, we describe them in the supplementary material.



\section{methodology}
\label{sec:method}
Our goal is to synthesis dense viewpoint in real time using our proposed lightweight view interpolation network based on the captured sparse viewpoint, and to perform tile-based dense view streaming strategy on the server side. Then, different clients can select the desired multi-view segment to download according to their instruction and switch smoothly among them, thus realising a one-to-many freeview system.

\subsection{View Interpolation Network}
\label{subsec:VINet}
Recently, optical flow has been widely used for video frame interpolation and achieved impressive
results,
however there has been few work on applying optical flow to view interpolation in the spatial domain. Inspired by {RIFE} \cite{huang2021rife}, a real-time video frame interpolation algorithm, we design a view interpolation network called VINet to synthesis the intermediate viewpoint between two cameras in real-time.

\paragraph{Network Architecture} The network architecture of our method is shown in Fig. \ref{VINet}. Let $\{I_{left},I_{right}\} \in \mathbb{R} ^{H\times W}$ be a pair of images which are captured by two adjacent cameras. The purpose of this network is to synthesis their intermediate viewpoint image $I_{mid}$ by simultaneously estimate the viewpoint switching flow and occlusion-aware mask, denoted by $\{F_{m\rightarrow l}, F_{m\rightarrow r}\}$ and $M$. Our solution can be expressed as
\begin{equation}
\begin{aligned}
I_{mid}(\boldsymbol u) = & M(\boldsymbol u) \cdot I_{left}{(\boldsymbol{u} + F_{m\rightarrow l}(\boldsymbol u))}\\
& + (1 - M(\boldsymbol u)) \cdot I_{right}(\boldsymbol u + F_{m\rightarrow r}(\boldsymbol u))
\end{aligned}
\end{equation}
where $I{(\boldsymbol{u} + F_{m\rightarrow l/r}(\boldsymbol u))}$ represents the pixel differentiablely warped from left/right viewpoint to intermediate viewpoint through the optical flow.

Three VIBlocks as shown in Fig. \ref{VINet} are used to estimate the flow and mask from coarse to fine, and a refinement network follows to improve the quality of the final synthesis image. The entire network pipeline is very efficient, achieving real-time requirements while maintaining visual quality.

\paragraph{Multi-Scale View Interpolation} In order to handle the viewpoint switching with large motion due to the long baseline between cameras, we implement a multi-scale flow estimation and view interpolation to achieve a coarse-to-fine effect. Specifically, we divide the network into three scales with gradually increasing resolution, which are handled by the corresponding three VIBlocks. The first stage estimates the rough flow and synthesizes the low resolution intermediate view which can handle the large motion. Then the next two stages refine the interpolated viewpoint by predicting the residual of the previous stage. This process can be expressed as follows.
\begin{equation}
\begin{aligned}
&F^{s}, M^{s} = F_{upscale} ^{s-1}, M_{upscale} ^{s-1} + Res^{s} \\
&I^{s} = M^{S} \odot I_{l\rightarrow m}^{s} + (1 - M^{s})\odot I_{r \rightarrow m}^{s}
\end{aligned}
\label{merge}
\end{equation}
where $\odot$ is an element wise multiplier, $I_{l\rightarrow m}$,$I_{r\rightarrow m}$ are the two warped images, and $s = 1, 2$. This setup allows our network to learn multiscale information, making it perform well in the quality of the reconstructed image.

\paragraph{Refinement Network} By weighted blending two warped images $I_{l\rightarrow m}$,$I_{r\rightarrow m}$ as expressed in \eqref{merge}, we can synthesis the final interpolated viewpoint image. However, this operation will cause a serious problem, i.e., a slight inaccuracy in optical flow estimation will result in severe blurring with this pixel-wise blending method.

Inspired by the previous work \cite{niklaus2018context}, we design a context-aware U-Net based residual refinement network to reduce sensitivity to optical flow accuracy. Specifically, we use a context extractor to extract rich contextual information from the original left and right viewpoint images, which are then fed into a U-Net with two pre-warped viewpoint images. Eventually, the refinement network will output the residual of the reconstructed image $\hat{I}_{m}$ to obtain a better visual effect.

\paragraph{Loss functions} We utilize a color-based loss function $\mathcal{L}_{Lap}$ and a feature-based loss function $\mathcal{L}_{F}$ to achieve perception-distortion tradeoff. By employing a multi-layer Laplace pyramid, $\mathcal{L}_{Lap}$ can reflect both local and global reconstruction information, which can be defined as follows.
\begin{equation}
\begin{aligned}
\mathcal{L}_{Lap} = \sum_{i=1}^{5}2^{i-1}\lVert L^{i}(\hat{I}) - L^{i}(I_{gt}) \rVert_{1}
\end{aligned}
\end{equation}
where $\hat{I},I_{gt}$ represent the synthesis image and its  corresponding ground truth, respectively. And $L^{i}$ represents the $i$-th layer of Laplace pyramid.

To achieve visually pleasing images, we also apply a loss $\mathcal{L}_{F}$ based on the higher-level features of the image and implement it in a multi-scale manner. 
\begin{equation}
\begin{aligned}
\mathcal{L}_{F} &= \lVert \phi(\hat{I}) - \phi(I_{gt}) \rVert_{1}    \\
\mathcal{L}_{ms} &= \sum_{i=0}^{2} w_i\cdot \mathcal{L}_{F}^{i}
\end{aligned}
\end{equation}
where $\phi$ denotes the feature extractor and $w_i$ represents the weights at different scales. Specifically, We find that VGG-19 \cite{simonyan2014very} loss performs best on our dataset and the configuration of our $w_i$ is [0.2, 0.3, 0.5].
 
\subsection{Streaming Strategy}
\label{subsec:streaming}
\begin{figure}
    \centering  
    \includegraphics[width=0.45\textwidth]{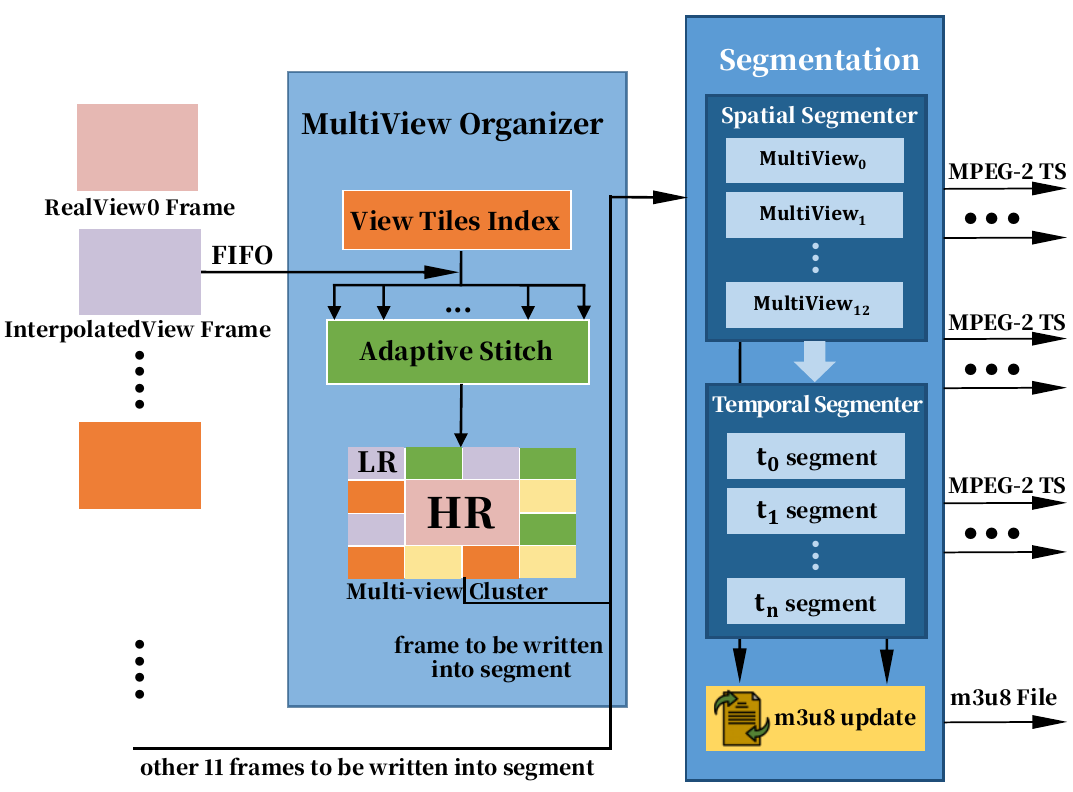} 
    \caption{Stream Strategy. Our multi-user oriented stream strategy can be divided into two main parts, namely multiview organizer and segmentation.} 
    \label{Stream}
\end{figure}

\setlength{\parskip}{0em} 
In order to meet the original intention of our system to be able to provide personalized services to multiple clients at the same time, we design a multi-user oriented streaming strategy as dipicted in Fig. \ref{Stream}. 

\paragraph{Multiview Organizer} Among the 
streaming schemes of 360° VR, tile-based method 
partitions the 360° video into a series of small tiles that can be transmitted and decoded independently. Only the tiles in FOV (Field of View) are transmitted with high resolution, while other tiles with low quality \cite{ahrar2021new}. 

Motivated by this method, we consider the real views captured by the cameras and the virtual views interpolated by our VINet as separate tiles. The high-resolution real view is stitched with the downscaled low-resolution interpolated viewpoint to form our organized multi-view cluster frame. Each of the 12 cameras has its own neighbouring interpolated virtual views, so that a total of 12 multi-view clusters are adaptively organized in parallel. The number of views within a cluster will be determined by the density of the view, ensuring a smooth switching experience. On the client side, the corresponding view is selected for decoding from the multi-view cluster, and in the case of a low-resolution interpolated view, it will perform super-resolution processing.

It is worth noting that, given the specificity of our tile-based frame, 
MCTS (motion-constrained tile set) of the codec should be enabled. It ensures that the encoding and decoding of the tiles will be completely independent, so that artifacts like mosaics will not appear.

\paragraph{Temporal-Spatial Segmentation} FVV allows users to gain an immersive experience by frequently switching viewpoint. This temporal and spatial switching characteristic places high demands on freeview video streaming. 

As shown in Fig. \ref{Stream}, we design a temporal-spatial segmentation method to cope with multi-user scenarios. It divides the whole multiple streams into a series of time domain and space(viewpoint) domain segments. Specifically, the 12 adaptively organized multi-view clusters output by MultiView Organizer form the segments on the spatial domain, and each of them is then split on the time order. This strategy allows our edge media server to serve as a media resource repository, where the server load is not related to the number of clients.

On the client side, users can switch from one view cluster to another over a period of time by downloading corresponding segments, and the tiles overlap between a portion of the view clusters ensures continuity of view switching.
\begin{table*}[t]
\begin{center}
\caption{Quantitative comparisons on three scenes of our dataset, and the inference time to synthesis a 720p frame. The numbers in
\textcolor{red}{\textbf{red}} indicate the best performance, while \textcolor{blue}{blue} indicate the second-best performance} \label{tab:Comparison}
\resizebox{\textwidth}{16mm}{
\begin{tabular}{c c|c|c| c|c|c| c|c|cc}
  \hline
  \multirow{2}{*}{\textbf{Methods}}  & 
  \multicolumn{3}{c|}{\textbf{Scene1} (baseline = 30 cm)} & 
  \multicolumn{3}{c|}{\textbf{Scene2} (baseline = 40 cm)} &
  \multicolumn{3}{c}{\textbf{Scene3} (baseline = 50 cm)} & \multirow{2}{*}{\textbf{Runtime(ms)}}\\ 
  \cline{2-10}
   & {PSNR}$\uparrow$ & {SSIM}$\uparrow$ & {LPIPS}$\downarrow$  
   & {PSNR}$\uparrow$ & {SSIM}$\uparrow$ & {LPIPS}$\downarrow$ 
   & {PSNR}$\uparrow$ & {SSIM}$\uparrow$ & {LPIPS}$\downarrow$\\
  
  \hline
  COLMAP\cite{schoenberger2016mvs} + VSS\cite{VSS} & 26.86  & 0.857  & 0.197 & 24.18  & 0.822  & 0.208 & 23.74  & 0.813  & 0.214  & 138315 + 28.72\\
  LLFF\cite{mildenhall2019local}          & 26.97  & 0.827  & 0.188 & 24.73  & 0.839  & 0.179 &  25.42  & 0.837  & 0.174  & 472
  \\
  Deep3DMask\cite{lin_deep_2021}         & 27.58  & 0.845  & 0.158 & 26.46  & 0.858  & 0.153 & 26.58  & 0.861  & 0.155  & 1267\\
  \hline
  Ours w/o Refine            & 32.92 & 0.924 & 0.059 & 30.66 & 0.897 & 0.084 & 29.95 & 0.879 & 0.092  &  \textcolor{red}{\textbf{6.27}}\\
    
    Ours - $\mathcal{L}_{Lap}$ & \textcolor{red}{\textbf{33.66}} & \textcolor{red}{\textbf{0.951}} & \textcolor{blue}{0.031} & 
    \textcolor{red}{\textbf{31.93}} & 
    \textcolor{red}{\textbf{0.928}} & 
    \textcolor{blue}{0.062} & 
    \textcolor{blue}{30.76} & 
    \textcolor{red}{\textbf{0.925}} & 
    \textcolor{blue}{0.078} &  
    \textcolor{blue}{12.85}\\
  Ours - $\mathcal{L}_{F}$   &
  \textcolor{blue}{33.15} & 
  \textcolor{blue}{0.938} & 
  \textcolor{red}{\textbf{0.008}} & 
  \textcolor{blue}{31.72} & 
  \textcolor{blue}{0.901} & 
  \textcolor{red}{\textbf{0.014}} &  
  \textcolor{red}{\textbf{30.84}} & 
  \textcolor{blue}{0.895} & 
  \textcolor{red}{\textbf{0.017}} &  \textcolor{blue}{12.85}\\
  \hline
\end{tabular}}
\end{center}
\end{table*}

\begin{figure*}
    \centering  
    \includegraphics[width=0.95\textwidth]{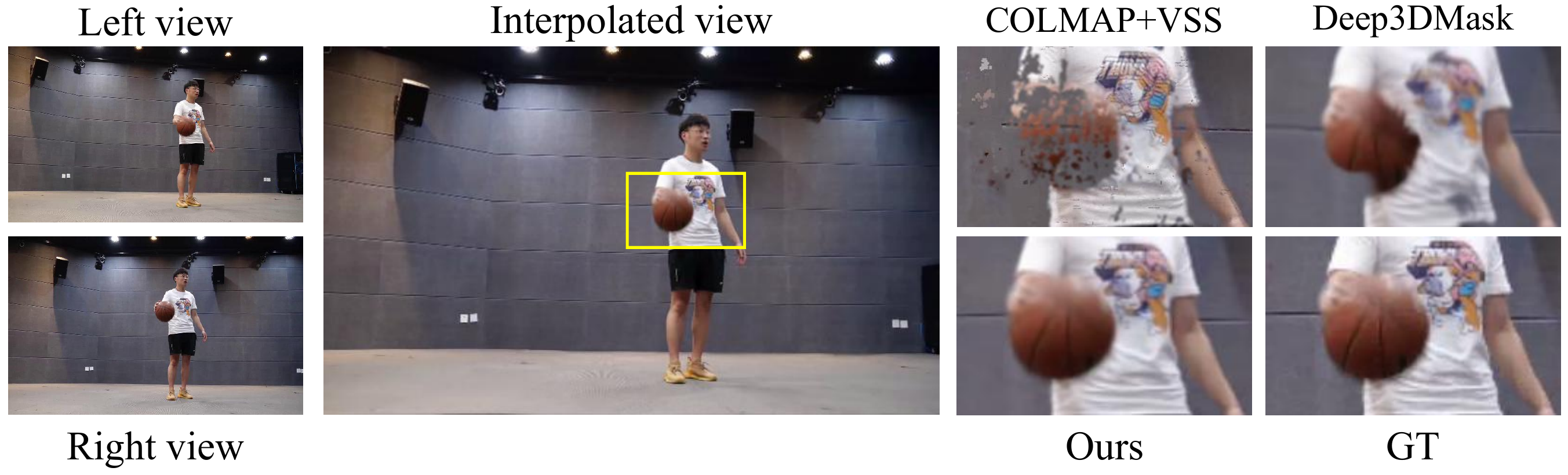}  
    \caption{Qualitative comparison on Scene3 validation set} 
    \label{comparison}
\end{figure*}
As for content delivery, HTTP Adaptive Streaming is a protocol that breaks
video contents into segments of the same duration and transmits these segments or sub-segments.  Authors of \cite{8326272} use HLS, which was developed by Apple, to realize a tile-based adaptive streaming system. Similar to this work, we utilize HLS to packet our encoded segments into the form of MPEG-2 TS files, and generate a live updated HLS-adapted m3u8 playlist. Strictly, the GOP size of encoded stream should correspond to one reasonable divisor of frame rate to make sure the first frame in segment file is an Intra frame.

Our temporal-spatial segmentation method enables different users to switch views by selecting the corresponding segment file to download as time changes. Moreover, the compatibility of streaming protocols allows our strategy to be plug-and-play easily.

\section{EXPERIMENTS AND RESULTS}
\subsection{Training Details}
\paragraph{Datasets} Vimeo90K \cite{xue2019video} triplet, which consists of 73,171 3-frame sequences with fixed resolution 448 × 256, is a widely used training
dataset in learning-based flow estimation and frame interpolation tasks. In our experiments, we first use it to pre-train our VINet, allowing it to learn an initial flow estimation weight. We also introduce a dataset with synchronized scenes captured by 12 cameras of our system.  Our cameras captured a total of 1000 frames in 60FPS with resolution 1920 × 1080 for each of the four scenes, and the 12 views in each frame can create 10 triplets, which together form our total dataset of 40,000 triplets. Our dataset will be released soon.

\paragraph{Implementation} We first train our network with $\mathcal{L}_{Lap}$ for 200 epochs with a learning rate decreases from $3e-4$ to $3e-5$ throughout the training process in a cosine annealing way. Then we use $\mathcal{L}_{F}$ to finetune 100 epochs to obtain a better visual perception. Our VINet is optimized by AdamW with weight decay $1e - 4$ in training stage. The whole training pipeline takes about 15 hours on two NVIDIA GeForce RTX 3090 GPUs, and our network is implemented using PyTorch. 
\subsection{Comparisons with  State-of-the-Art works}
\setlength{\parskip}{0em} 
We compare our VINet to several state-of-the-art view synthesis algorithms available in terms of image quality and running speed. These methods cover the DIBR based method VSS \cite{VSS}, which is most commonly used in industry, and the depth map is obtained using COLMAP\cite{schoenberger2016mvs}. MPI-based methods such as Deep3DMask\cite{lin_deep_2021}, which uses a 3D mask to obtain a good result, and LLFF\cite{mildenhall2019local}, which performs very well in terms of inference speed, have also been compared. 

\begin{table*}[t]
\begin{center}
\caption{Latency performance per frame on 1000 iterations} \label{latency}
{
\begin{tabular}{|c|cccc|cc|}
  \hline
  \multirow{2}{*}{\diagbox{\textbf{Latency}}{\textbf{Module}}} & 
  \multicolumn{4}{c|}{\textbf{Server}} & 
  \multicolumn{2}{c|}{\textbf{Clients}} \\ 
  \cline{2-7}
   & view interpolation & adaptive stitch & encoder & schedule & PC & Phone
  \\
  \hline
  {Maximum} & 17.89ms & 2.08ms & 9.62ms & 8.97ms & 10.72ms & 19.63ms \\
  {Minimum} & 11.24ms & 0.98ms & 4.27ms & 3.21ms & 2.16ms & 9.81ms \\
  {Average} & 12.85ms & 1.27ms & 5.62ms & 5.45ms & 4.62ms & 11.43ms \\
  \hline
\end{tabular}}
\end{center}
\end{table*}

We consider the left and right viewpoint images as input and the intermediate view as ground truth, and measure the PSNR, SSIM and LPIPS between the synthesized image and the GT. We select three scenes from our dataset with different camera baselines for comparison. From the quantitative results of the three scenes in Table \ref{tab:Comparison}, we can find that our VINet outperforms the current SOTA methods in terms of quality of the synthesized images. Further ablation experiments also show that our refinement network and perception-distortion tradeoff contribute some gains to our results. Some visual results are shown in Fig. \ref{comparison}. Please also refer to the supplementary video demo to examine more visual quality of our results.

Since only intermediate view between two cameras can be interpolated at a time, we can recursively stack VINets through the pipe to obtain a sufficiently dense views. Specifically, a stack of $n$ stages will obtain a total of $2^n - 1$ views between the two cameras. The 4-stage network generates a total of 15 views between the two cameras, which is sufficient to bring the viewer a smooth switching experience. A visual result on dense view generation is in the supplementary material.
\subsection{System  Evaluation}
Our whole system is written in C++ and our VINet is deployed using LibTorch. In order to obtain an overview of the performance of our system, we test the latency of each critical kernel module separately and the end to end latency is shown in Table \ref{latency}. All kernel modules in our system are highly parallel, so the frame rate on the server side is limited only by the most time-consuming module, which is about 13ms for view interpolation, thus ensuring that it can achieve up to 77 FPS. 


\section{conclusion}
In this paper, we build an end-to-end real-time free-viewpoint system with an efficient view interpolation network VINet, being integrated into it. Our VINet is capable of interpolating the dense view between two adjacent cameras in real time. Moreover, a multi-user oriented dense view streaming strategy is proposed. The whole system is able to provide high-quality live freeview services to a theoretically arbitrary number of users using a sparse distribution of cameras. Our future work includes multi-view encoding to reduce the amount of data transmitted, and reference-based super-resolution to improve the quality of low-resolution interpolated views on the client side.

%

\bibliographystyle{IEEEbib}
\bibliography{ICME2022}
\clearpage
\section{Supplementary Materia}
\subsection{System Configuration and Scheduling}
\subsubsection{Hardware Configuration}
\paragraph{Capturer} 12 different Panasonic LUMIX BGH1 cameras, four capture cards and four capture servers make up our capturer cluster. All cameras are arranged in a form of circular arc, with a view angle difference of about 5 degrees between each. Each of 12 different camera is able to capture 4K live video at 60FPS and synchronise with other cameras via hardware. Capture server with a capture card connected controls three cameras separately, taking charge of their video encoding and sinking task.

\paragraph{Media Server} The Media Edge Server is equipped with four NVIDIA RTX A6000 GPUs. There is no limit to the number of NVENC sessions on this graphics card, meaning that it can simultaneously encode multiple videos in real-time with hardware as long as the computing power is sufficient, which is perfect for our multi-camera setup. In our system, one graphics card encodes video from three cameras. 

It also has two Intel Xeon Gold 6240 CPU and 256GB RAM with an SSD for data storage. To ensure real-time performance, we run most of the computationally complex modules on the GPUs, including our VINet inference, adaptive multi-view stitching, and encoding of multi-view clusters.

\paragraph{Clients} In our experimental configuration, Clients are represented by a PC and a cell phone. As for the PC client, an Intel Core i9 CPU and one Nvidia RTX 3070 GPU are equipped for video decoding. This is a regular consumer-level configuration, and although we use the latest graphics cards, it performs similarly to previous generations of chips, so that even lower-configured clients can access the service. 

In addition, a Samsung Galaxy S21 is used as our mobile phone client. As well, since the client accessing our system does not require any rendering operations, a low-cost phone can also be competent.It should be noted that although we use only two clients, an arbitrary number of clients can access our system and receive personalized interaction services. 
\subsubsection{Schedule Scheme}
As described in the main paper, We adopt a 4\textbf{U} resource model to schedule our system. Specifically, \textbf{U}nified Data Pack denoted by $p$ contains data type, data size and a block of binary data.
$$p = [type \quad size  \quad data]$$

It is designed to cope with the diversity of data formats in video systems, thus alleviating the burden of data management. The data of each frame is represented by an \textbf{U}nified Data Pack.  The data type distinguishes whether the data of the current frame is stored in the CPU or in the GPU, while the data size is used to cope with different frame resolution and data formats(YUV420/RGB).
\begin{figure}
    \centering  
    \includegraphics[width=0.48\textwidth]{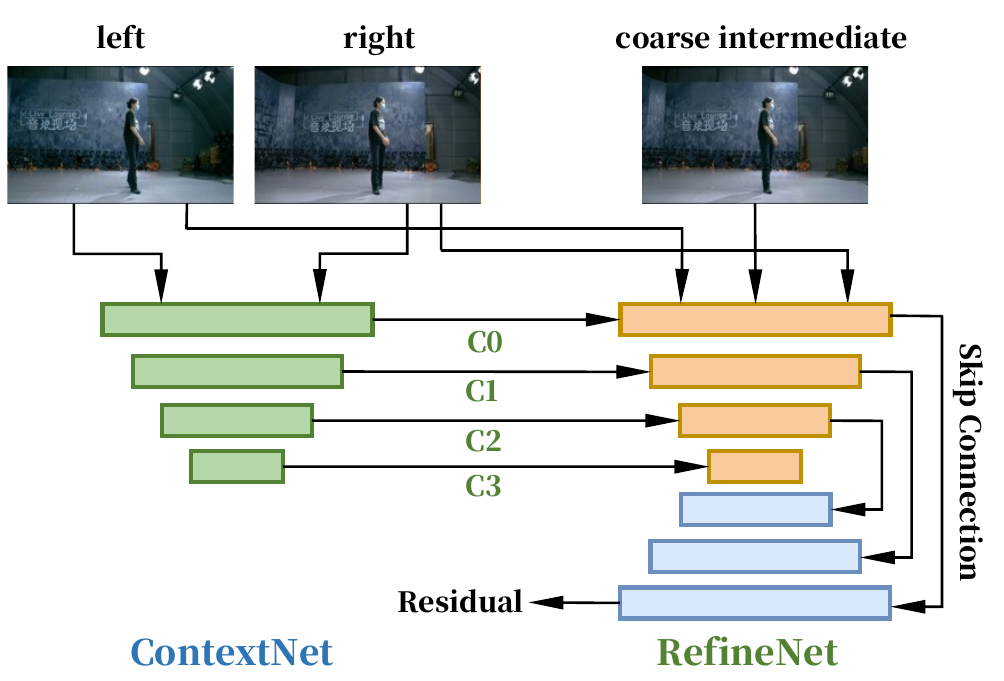}  
    \caption{The structure of our context-aware RefineNet} 
    \label{refinenet}
\end{figure}
\textbf{U}niversal data channel with a first-in-first-out(FIFO) structure to pass data packets between different task modules, which can be expressed as

$$c = [type \quad count  \quad FIFO]$$

The FIFO offers two operations, PUT and GET, for packet communication, and guarantees data security through mutexes. And a globally accessible data interface is equipped with Universal data channel to simplify data communication in video system. Similar to \textbf{U}nified Data Pack, the type label  indicates whether the packet is from the CPU or GPU and the count represents the number of available data packets. An \textbf{U}niversal data channel can be  registered by diverse tasks. Moreover, the one-in multi-out function ensures the reusability of the data channel.

The other 2\textbf{U} denote the \textbf{U}nitary Task Kernel, to cope with dynamic reconfigurations and the  \textbf{U}ltimate Workflow Pipeline, which is used to increase system  elasticity.

\subsection{Details of Our VINet}
Our VINet architecture consists mainly of three efficient VIBlocks and a context-aware RefineNet. Among them, VIBlocks are designed to estimate the optical flow at three different scales, and the latter stage is for estimating the residual of the previous stage, thus achieving a coarse-to-fine effect. RefineNet further repairs the intermediate view obtained by spatial warping, and its structure is shown in Fig. \ref{refinenet}. It extracts the rich contextual information of the left and right views with a contextnet, and then estimates the residuals by a Unet-based network to obtain a more satisfactory result. The details of each layer of our whole VINet are listed in Table \ref{tab:layer details}.

\begin{table}[t]
\begin{center}
\caption{ Details of each layer in our VINet. H,W represents the ratio to the original input size and $i=(0,1,2)$ denotes the $i$-th level of VINet.} 
\label{tab:layer details}
\resizebox{0.48\textwidth}{47mm}
{
\begin{tabular}{c ccc}
  \hline
  {\textbf{SubNet}}  & Operation &  H,W &  Out Channels \\
  \hline
  \multirow{5}{*}{{VIBlock$_i$}}
                     & Conv2d(stride 2)-RELU &  2$^i$/8 &   128   \\
                     & Conv2d(stride 2)-RELU &  2$^i$/16 &  256   \\
                     & 6 $\times$ (Conv2d-RELU) &  2$^i$/16 &   256   \\
                     & ConvTranspose2d(stride 2)-RELU &  2$^i$/8 &  128 \\
                     & ConvTranspose2d &  2$^i$/4 &  5 \\
  
  \hline
  \multirow{8}{*}{{ContextNet}}
                     & Conv2d(stride 2)-RELU &  1/2  &  16   \\
                     & Conv2d-RELU           &  1/2  &  16   \\
                     & Conv2d(stride 2)-RELU &  1/4  &  32   \\
                     & Conv2d-RELU           &  1/4  &  32   \\
                     & Conv2d(stride 2)-RELU &  1/8  &  64   \\
                     & Conv2d-RELU           &  1/8  &  64   \\
                     & Conv2d(stride 2)-RELU &  1/16 &  128   \\
                     & Conv2d-RELU           &  1/16 &  128   \\

  \hline
  \multirow{13}{*}{{RefineNet}}
                     & Conv2d(stride 2)-RELU &  1/2  &  32   \\
                     & Conv2d-RELU           &  1/2  &  32   \\
                     & Conv2d(stride 2)-RELU &  1/4  &  64   \\
                     & Conv2d-RELU           &  1/4  &  64   \\
                     & Conv2d(stride 2)-RELU &  1/8  &  128   \\
                     & Conv2d-RELU           &  1/8  &  128   \\
                     & Conv2d(stride 2)-RELU &  1/16 &  256   \\
                     & Conv2d-RELU           &  1/16 &  256   \\
                     & ConvTranspose2d(stride 2)-RELU &  1/8 &  128 \\
                     & ConvTranspose2d(stride 2)-RELU &  1/4 &  64 \\
                     & ConvTranspose2d(stride 2)-RELU &  1/2 &  32 \\
                     & ConvTranspose2d(stride 2)-RELU &  1   &  16 \\
                     & Conv2d-Sigmoid                 &  1   &  3   \\
\end{tabular}}
\end{center}
\end{table}

\begin{figure*}
    \centering  
    \includegraphics[width=\textwidth]{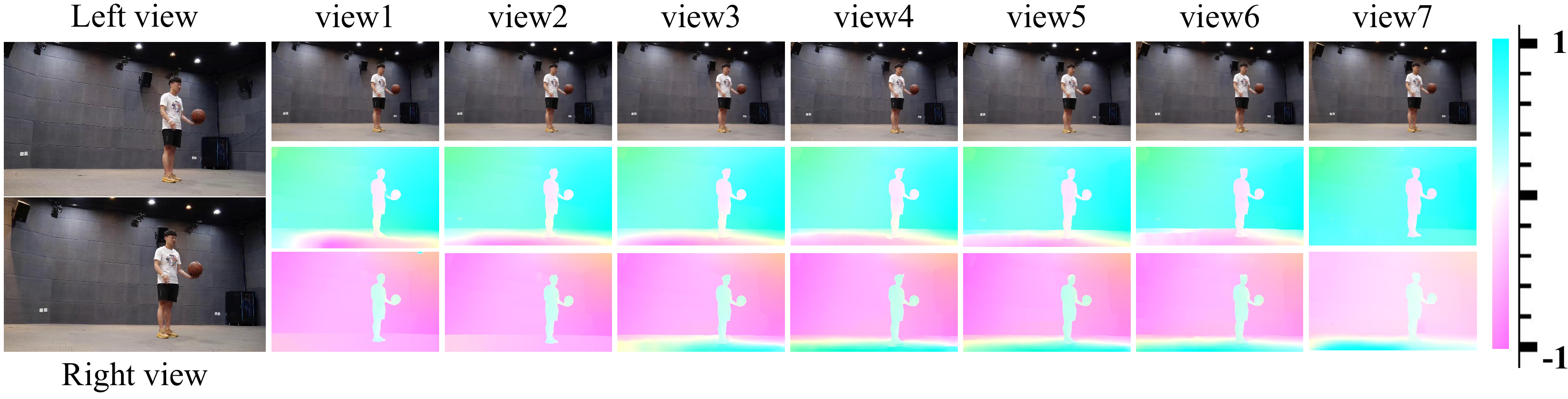}  
    \caption{A result of a three-stage VINets on Scene3 validation set. The left-most column are the input two views. In the remaining columns, the first row are the 7 interpolated views. The remaining two rows represent the optical flow between the interpolated view and the left/right view, i.e. the position offset.} 
    \label{fig:multiview}
\end{figure*}
    
\subsection{Dense view generation}
As mentioned in the main paper, we can recursively stack VINets through the pipe to obtain a sufficiently dense views.  Specifically, given the images of two adjacent cameras$I_{left}, I_{right}$, our VINet can synthesize intermediate view image $I_{mid}$. Then, the second stage VINet can generate a quarter-position view through $I_{left}, I_{mid}$. We repeat the above operation, then a stack of $n$ stages will obtain a total of $2^n - 1$ views between the two cameras. We interpolate 7 views and obtain a total of 9 views between the two cameras by stacking a three-stage VINet.  As shown in Fig. \ref{fig:multiview}, the quality of these interpolated view is satisfying. To more visually represent the positional offsets between the interpolated view and the left/right view, we visualize the optical flow between each interpolated view and the left/right view in Fig. \ref{fig:multiview}. The positional offsets are expressed as pixel offsets which can be represented as two-dimensional vectors, and we normalize them. 

For instance, the second row in Fig. \ref{fig:multiview} represents the optical flow between the interpolated image and the left image, which has a positive pixel offset and thus exhibits a blue color. On the contrary, the third row represents the optical flow between the interpolated image and the right image, which has a negative pixel offset thus appears red. The darker the color, the greater the offset. The corresponding scale is given on the rightmost side in Fig. \ref{fig:multiview}.

In our system, a total of 4 stages of VINet are deployed, which means that a total of 177 views can be interpolated from our 12 cameras. As for our multi-view cluster, we chose 16 views on each side of the camera for stitching, so that the user can switch between 33 views within the duration of one segment. Our experiments demonstrate that with such a dense number of views, the interaction is very smooth while the view switching latency is very low.

\subsection{Streaming Client}
\begin{figure*}
    \centering  
    \includegraphics[width=0.7\textwidth]{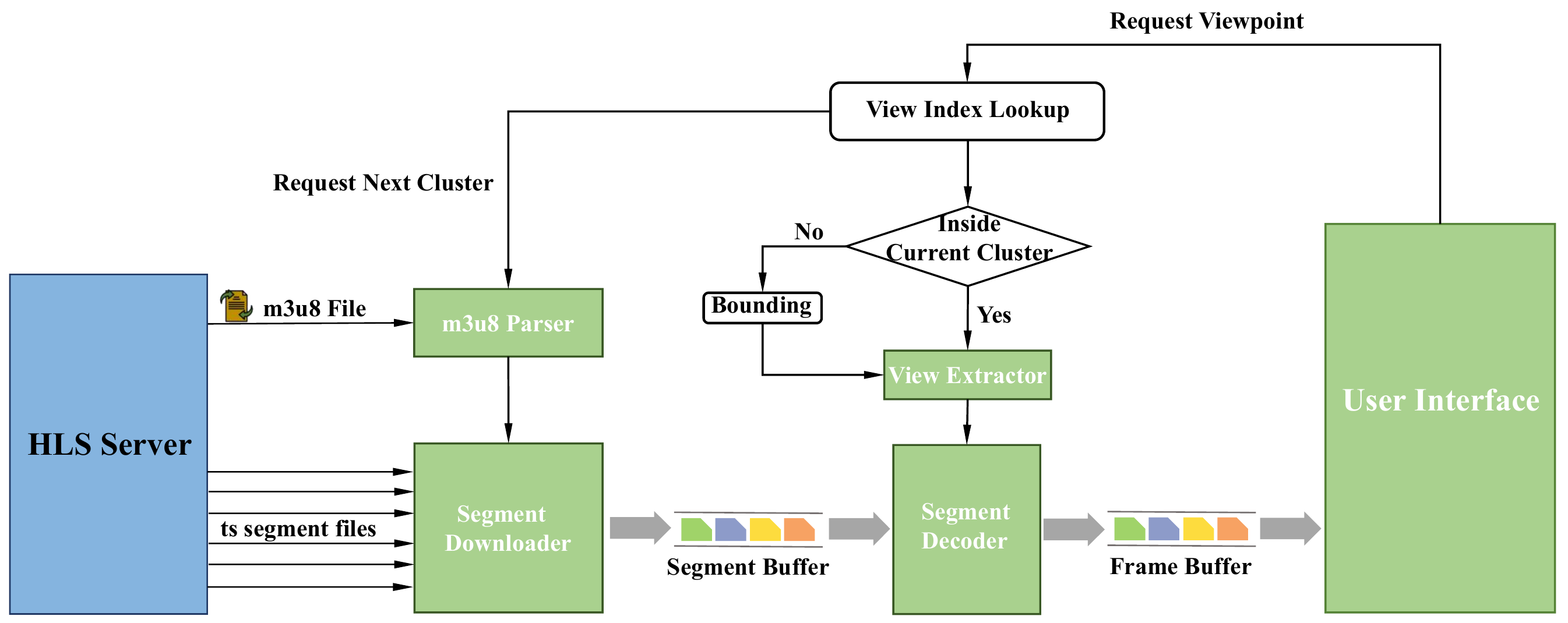}  
    \caption{The components of our streaming client.} 
    \label{fig:client}
\end{figure*}

 As illustrated in Fig. \ref{fig:client}, the streaming client of our system consists of a number of modules, including \textit{m3u8 Parser, Segment Downloader, View Extractor, Segment Decoder} and \textit{User Interface}. The tasks executed in these modules all impose only a minimal computational load, so that some thin clients are capable of handling them, such as mobile devices.
View Extractor is designed to extract the bitstream of the current viewpoint tile. When the user interactively requests the corresponding viewpoint, an operation to find the corresponding view index from the global lookup table is executed. If the desired viewpoint is in the current multi-view cluster, View Extractor extracts the bitstream of the corresponding viewpoint tile and decodes it by the Segment Decoder. Otherwise, the view switching is restricted to the boundary of the current multi-view cluster and View Extractor extracts the information of the boundary viewpoint.
 
The user's instruction directly determines the next segment to be downloaded. Specifically, the user's desired viewpoint is immediately selected in the current segment. As the viewpoint is switched, the m3u8 parser adaptively selects the closest segment from the m3u8 file, and then the Segment Downloader downloads the corresponding ts file. The Segment Decoder is responsible for decoding the extracted bitstream into raw YUV and sending it to the frame buffer, which is finally played in the user interface.

The switching latency on our client side is very low, since all switching operations are performed locally and there is no network response latency. Additionally, the heaviest computational load of all client-side tasks is decoding, a task that many devices can easily handle, so that our system can provide services to low-end, heterogeneous client terminals.

\end{document}